\def \AAP #1 #2 {{\em Astron. Astrophys.\/} {\bf #1}, #2}
\def \AAL #1 #2 {{\em Astron. Astrophys. Lett.\/} {\bf #1}, L#2}
\def \AAR #1 #2 {{\em Astron. Astrophys. Rev.\/} {\bf #1}, #2}
\def \AAS #1 #2 {{\em Astron. Astrophys. Suppl. Ser.\/} {\bf #1}, #2}
\def \AJ #1 #2 {{\em Astron. J.\/} {\bf #1}, #2}
\def \ANNREV #1 #2 {{\em Ann. Rev. Astron. Astrophys.\/} {\bf #1}, #2}
\def \APJ #1 #2 {{\em Astrophys. J.\/} {\bf #1}, #2}
\def \APJL #1 #2 {{\em Astrophys. J. Lett.\/} {\bf #1}, L#2}
\def \APJS #1 #2 {{\em Astrophys. J. Suppl.\/} {\bf #1}, #2}
\def \APSS #1 #2 {{\em Astrophys. Space Sci.\/} {\bf #1}, #2}
\def \ASR #1 #2 {{\em Adv. Space Res.\/} {\bf #1}, #2}
\def \BAIC #1 #2 {{\em Bull. Astron. Inst. Czechosl.\/} {\bf #1}, #2}
\def \JSQRT #1 #2 {{\em J. Quant. Spectrosc. Radiat. Transfer\/} {\bf #1}, #2}
\def \MN #1 #2 {{\em Mon. Not. R. Astr. Soc.\/} {\bf #1}, #2}
\def \MEM #1 #2 {{\em Mem. R. Astr. Soc.\/} {\bf #1}, #2}
\def \PLR #1 #2 {{\em Phys. Lett. Rev.\/} {\bf #1}, #2}
\def \PASJ #1 #2 {{\em Publ. Astron. Soc. Japan\/} {\bf #1}, #2}
\def \PASP #1 #2 {{\em Publ. Astr. Soc. Pacific\/} {\bf #1}, #2}
\def \NAT #1 #2 {{\em Nature\/} {\bf #1}, #2}
\def \SAIT #1 #2 {{\em Mem.\ Soc.\ Astron.\ It.\/} {\bf #1}, #2}
\def \MESS #1 #2 {{\em The Messenger\/} {\bf #1}, #2}
\def \ASTRNACH #1 #2 {{\em Astron. Nach.\/} {\bf #1}, #2}

\documentstyle{BSAXPwork}
\input epsf.sty
\begin{opening}
\title{Bulk viscosity, r-modes, and the early evolution of neutron stars}
\author{Andreas Reisenegger
\& Axel Bona\v ci\'c
}
\institute{
Departamento de Astronom\'\i a y Astrof\'\i sica, Facultad de
F\'\i sica, Pontificia Universidad Cat\'olica de Chile, Casilla
306, Santiago 22, Chile; areisene,abonacic@astro.puc.cl}
\date{} 
\end{opening}

\begin{document}

\oddpagefooter{}{}{} 
\evenpagefooter{}{}{} 
\medskip  

\begin{abstract} 
We discuss the effect of nonlinear bulk viscosity and the
associated reheating on the evolution of newly born, rapidly
rotating neutron stars with r-modes destabilized through
the Chandrasekhar-Friedman-Schutz (CFS) mechanism.
Bulk viscosity in these stars is due to
the adjustment of the relative abundances of different particle
species as the density of a fluid element is perturbed. It becomes
nonlinear when the chemical potential difference $\delta\mu$,
measuring the chemical imbalance in the fluid element, becomes
larger than the temperature $T$, which is generally much smaller
than the Fermi energy. From this scale on, the bulk viscosity
increases much faster with $\delta\mu$ than predicted by the
usual, linear approximation. This provides a potential saturation mechanism
for stellar oscillation modes at a small to moderate amplitude. In
addition, bulk viscosity dissipates energy, which can lead to
neutrino emission, reheating of the star, or both.
This is the first study to explicitly consider these effects in the
evolution of the r-mode instability.
For stars with little or no hyperon bulk viscosity, these effects are
not strong
enough to prevent the r-modes from growing to amplitudes $\alpha\sim 1$
or higher, so other saturation mechanisms will probably set in earlier.
The reheating effect makes spin-down occur at a higher temperature than
would otherwise be the case, in this way possibly avoiding complications
associated with a solid crust or a core superfluid. On the other hand,
stars with a substantial hyperon bulk viscosity and a moderate magnetic
field saturate their mode amplitude at a low value, which makes them
gravitational radiators for hundreds of years, while they lose angular
momentum through gravitational waves and magnetic braking.

\end{abstract}

\medskip

\section{Introduction}

R-modes in inviscid, rotating, relativistic stars (in which the Coriolis force acts as
the restoring force to produce periodic fluid motions) are
generically unstable to the emission of gravitational waves
(Andersson 1998; Friedman \& Morsink 1998), whose
back-reaction produces a growth of the modes by the mechanism
first studied by Chandrasekhar (1970) and by Friedman \& Schutz
(1978a,b; ``CFS instability'').

In neutron stars at low temperatures $T$, the long mean free
path of the fluid particles allows rapid momentum exchange over
macroscopic lengthscales (shear viscosity), impeding the growth
of the modes (Lindblom et al. 1998).
At high $T$, the successive compression and expansion of a fluid
element (a small effect for r-modes, unless the stellar rotation
rate, $\Omega$, is extremely fast; e.g., Lindblom \& Owen 2002)
leads to a departure from the
chemical equilibrium conditions, inducing reactions (typically
weak interactions) which tend to restore the equilibrium state.
Since these reactions occur out of equilibrium, they constitute
irreversible processes that dissipate energy and create entropy,
extracting mechanical energy from the oscillation modes while
increasing the temperature of the fluid and/or leading to the
emission of neutrinos from the star.

The competition between the destabilizing effect of the
gravitational radiation reaction, which is strongly dependent on
$\Omega$, and
the damping effect of the temperature-dependent shear and bulk
viscosity generically defines one or two ``instability regions''
for low-amplitude r-modes at high $\Omega$ and intermediate $T$,
as shown in Fig. 1. In these regions, the mode amplitude grows
exponentially, until it is choked by cooling or feeds back on
$\Omega$ and $T$ (and itself) through heating or
gravitational-radiation torque.

\begin{figure}
\epsfysize=8cm 
\hspace{1.0cm} \vspace{0.0cm}
 \epsfbox{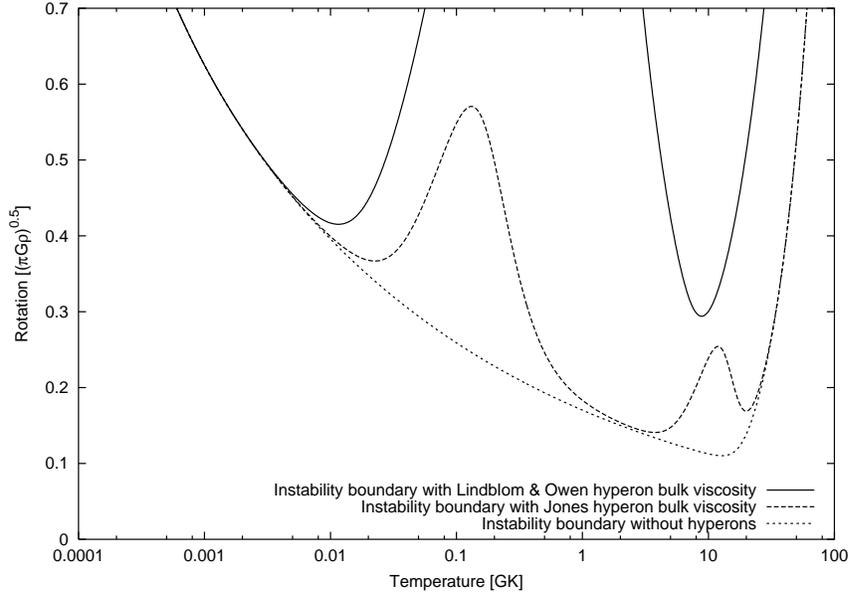}
\caption[h]{Instability boundaries obtained for different hyperon bulk
viscosity calculations (Jones 2001; Lindblom \& Owen 2002), and for a
non-hyperonic neutron star, all with the shear viscosity calculated
by Levin \& Ushomirsky (2001b) for a neutron star with an elastic
crust. The unstable region is that above each instability boundary.}
\end{figure}

Studies done so far of the evolution of young neutron stars under
this instability (e.g., Owen et al. 1998; Lindblom \& Owen 2002)
include amplification and damping of r-modes by the
above-mentioned effects in the linear regime, spin-down due to
gravitational waves, and passive cooling, ignoring the nonlinear
terms and the heating associated with the bulk viscosity. These
effects are included in the present work, in addition to a
probably more realistic treatment of angular momentum loss,
motivated by the model case studied by Levin \& Ushomirsky (2001a).

\section{Nonlinear bulk viscosity and reheating}

In the simplest case, namely neutron star matter composed only of
neutrons ($n$), protons ($p$), and electrons ($e$), the
equilibrium state is set by $\mu_n=\mu_p+\mu_e$, where $\mu_i$ is
the chemical potential of particle species $i$. In the simplest
approximation, we consider non-interacting particles, of which
the neutrons and protons are non-relativistic, satisfying
$\mu_i\approx p_{Fi}^2/(2m_i)\propto n_i^{2/3}$, where $p_{Fi}$ and $n_i$
are the respective Fermi momenta and number densities, and
the electrons are extremely relativistic, with
$\mu_e\approx p_{Fe}c\propto n_e^{1/3}$. If the total density $\rho$
is perturbed by an amount $\delta\rho$, at constant (charge-neutral)
composition, the equilibrium state is perturbed by
$\delta\mu\equiv\mu_p+\mu_e-\mu_n\approx-\mu_e\delta\rho/(3\rho).$

The equilibrium state can be restored by the ``modified Urca reactions''
(hereafter {\it mUrca}), $n+N\to p+N+e+\bar\nu_e$ and $p+N+e\to n+N+\nu_e$,
where $\nu_e$ and $\bar\nu_e$ are the electron neutrino
and antineutrino, and $N$ is a ``bystander nucleon'' whose
identity is not changed, but whose function is to absorb momentum
in order to allow for overall momentum conservation in the
reaction, given that the neutron Fermi momentum is much larger
than that of all the other particles involved.

The net rate $\Gamma_m$ of these reactions (net lepton number
emitted per unit volume per unit time) is limited by the available
phase space, increasing both with temperature, $T$, and with
$\delta\mu$. The associated energy dissipation rate is

\begin{equation}
\Gamma_m(\delta\mu,T)\delta\mu=\epsilon_m(0,T)
{14680u^2+7560u^4+840u^6+24u^8\over 11513},
\end{equation}

\noindent where $u\equiv\delta\mu/(\pi T)$, and the neutrino
emissivity (energy emitted in the form of neutrinos and
antineutrinos per unit time per unit volume) takes a similar form,

\begin{equation}
\epsilon_m(\delta\mu,T)=\epsilon_m(0,T)
\left(1+{22020u^2+5670u^4+420u^6+9u^8\over 11513}\right),
\end{equation}

\noindent with the equilibrium value $\epsilon_m(0,T)\propto T^8$
(see Reisenegger 1995). Studies of neutron star evolution so far
have considered only the lowest-order term in $u$ in each of these
expressions, which yield the linear bulk viscosity and equilibrium
neutrino cooling of the star, respectively. Here, we consider the
full expression for both. Equation (1) shows that, at high
amplitudes, the dissipation increases much more strongly than predicted
by linear theory, potentially providing a saturation mechanism
that impedes the unlimited growth of the amplitude. In addition,
we note that the net heating of the neutron star is given by the
difference $\Gamma_m\delta\mu-\epsilon_m$, a fourth-order
polynomial in $u^2$ whose zeroth and first-order terms are
negative whereas the three higher powers are positive. Thus, for
small-amplitude oscillations, the star undergoes net cooling
(somewhat faster than in equilibrium at the same $T$), whereas at
high amplitudes it is heated more and more strongly.

In the inner core of a neutron star, additional particles may
appear, such as the (baryonic) $\Lambda^0$ and $\Sigma^-$
hyperons. Their presence changes the equilibrium
abundances, increasing the proton-to-neutron
ratio and probably allowing the much faster ``direct Urca
reactions'' (hereafter {\it dUrca}), $n\to p+e+\bar\nu_e$ and
$p+e\to n+\nu_e$, to take place, conserving
momentum without the need of a ``bystander particle''. For these, the
quantitative details are slightly different from the
{\it mUrca} case (see Reisenegger 1995 for the explicit expressions),
but the qualitative properties discussed above still hold.

In addition, the hyperons give rise to the additional reactions
$\Sigma^-+p\leftrightarrow n+n$ and $\Lambda^0+p\leftrightarrow
p+n$, which contribute to the bulk viscosity,
substantially reducing the unstable region in the $\Omega-T$ plane,
probably to two small and disjoint ``windows'' (see Fig. 1 and
Lindblom \& Owen 2002).
These reactions emit no neutrinos,
and therefore always
have a heating effect. The rates of these reactions have only been
calculated in the small-amplitude (``linear'') limit (Jones
2001; Lindblom \& Owen 2002), but we extrapolate them into the
nonlinear regime by using the expressions obtained by Madsen
(1992) for the analogous processes in quark matter,
$u+d\leftrightarrow s+u$.

\section{Angular momentum loss and neutron star evolution}

The evolution of the star is modeled by
coupled first-order differential equations for $\Omega$, $T$, and
the dimensionless mode amplitude $\alpha$. The evolution of $T$
is determined by the heating and cooling processes discussed in
the previous section, and that of $\alpha$ by the mentioned
competition between the destabilizing gravitational radiation
reaction and the stabilizing viscous effects. The evolution of
$\Omega$ deserves some additional comments.

In the instability conditions
derived by Friedman \& Schutz (1978b), a crucial role is played
by the so-called ``canonical angular momentum'' associated with
the mode, a quadratic functional of the displacement field which
is related (though possibly not identical) to the physical angular
momentum of the mode. In previous studies of the evolution of neutron
stars under this instability (e.g., Owen et al. 1998),
it was assumed that the total angular momentum of the star is the sum
of a term corresponding to a rigid rotation, $I\Omega$, where $I$ is
the star's (constant) moment of inertia, plus the canonical angular
momentum of the mode. Furthermore, it was assumed that only the latter
term is modified by the gravitational radiation reaction force, whereas
the stellar rotation rate can be changed only by viscous dissipation
of the modes, which would transfer angular momentum from one term to the
other.

For one specific model problem (a thin spherical shell), Levin \& Ushomirsky
(2001a) were able to show that these assumptions are incorrect. For this
case, the r-modes have no physical angular momentum, i.e.,
do not contribute to the total angular momentum of the star, which is exactly
$I\Omega$. Furthermore, $I\Omega$ is changed by the gravitational waves, as
there is no other angular momentum available for them to carry away.
There is no proof that these
results can be extrapolated to the much more complicated case of a complete
neutron star. However, given that the previously made assumptions are rather
special and have been disproved in the only available model case, we
choose to base our evolutionary model on the results of Levin \& Ushomirsky,
taking the total angular momentum to be given by $I\Omega$ and its derivative
to be equal to minus the angular momentum carried away by the gravitational
waves.

\section{Results}

We have tracked the evolution of a young neutron star for three different
scenarios, which are determined by the instability boundaries
shown in Fig. 1. Our numerical simulations start with the same initial
conditions as in Owen et al. (1998), i.e., an initial temperture
$T_0=10^{11}\rm{K}$ and an initial rotation rate
$\Omega_0=\frac{2}{3}\sqrt{\pi G\bar{\rho}}$,
about the maximum rotation the star can sustain without disrupting due
to the centrifugal force, where $\bar{\rho}$ is the mean density of the star.

\begin{figure}
\epsfysize=8cm 
\hspace{1.0cm} \vspace{0.0cm}
 \epsfbox{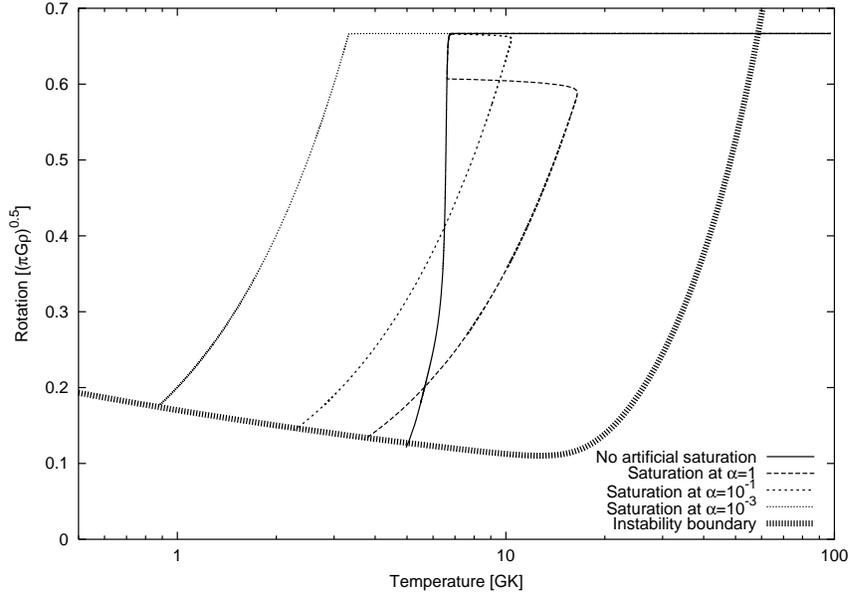}
\caption[h]{Evolution of a non-hyperonic neutron star. The solid line
corresponds to the case without artificial saturation. The segmented
lines represent the evolution with artificial saturation of the r-modes at
different amplitudes $\alpha$, as indicated.}
\end{figure}

Our first scenario is a neutron star composed only of neutrons, protons and
electrons, where its bulk viscosity is caused by {\it mUrca} reactions
(Fig. 2).
Non-linear bulk viscosity is unable to saturate the r-modes
on its own, so they can grow to amplitudes $\alpha\geq1$,
spinning down the star within an hour. Reheating is weak compared to neutrino
emission, leading only to
a delay in the cooling of the star, but not to a net temperature increase.
Since other saturation mechanisms can be active (Arras et al. 2002),
we have saturated the r-modes artificially (as in
Owen et al. 1998) by an ad hoc damping force whose work is directly dissipated
into heat. This leads to a rapid and substantial temperature increase,
depending on the saturation amplitude,
until the star is hot enough for neutrino cooling to become stronger. For
saturation amplitudes $\alpha<10^{-2}$, as for the unsaturated case,
the heat gain is not enough to raise
the temperature and only delays the cooling.
However, the spin-down timescale
becomes much longer, $t_{sd}\sim\alpha^{-2}$ hours.
This mechanism can in principle explain
the inexistence of very rapidly rotating, young neutron stars.
However, with our current understanding of the microphysics, the instability
boundary restricts the final rotation periods (after spin-down through this
mechanism) only to $P>5$ ms, not to the observed $P>15$ ms.

In our second scenario (not plotted), we consider the presence of hyperons in the inner
core of the neutron star, contributing the relatively low bulk viscosity
estimated by Jones (2001) and allowing {\it dUrca} processes to take place.
The evolution is similar to our
first scenario, but happens at lower temperatures ($\sim 10^9\rm{K}$),
allowing some temperature increase to occur once the mode amplitude is large.

\begin{figure}
\epsfysize=8cm 
\hspace{1.0cm} \vspace{0.0cm}
 \epsfbox{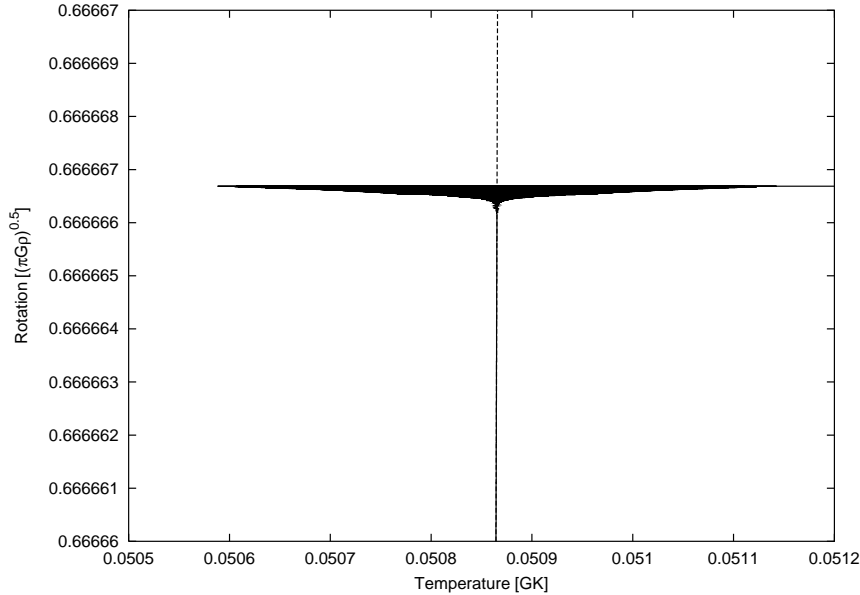}
\caption[h]{Evolution of a neutron star with large hyperon bulk
viscosity (solid line) when it gets to the second unstable region,
whose boundary is shown for reference (dashed line).}
\end{figure}

Our third scenario is basically the same as the second one, but the
contribution of hyperon bulk viscosity is taken to be large (as suggested by Lindblom
\& Owen 2002), so as to split the unstable
region in two (Fig. 1). Due to the rapid {\it dUrca} cooling, the high-temperature
unstable region is passed through in seconds, and initially small-amplitude
r-modes do not have enough time to grow significantly. It
takes about a year for the neutron star to cool down to the second unstable region,
which is avoided due to the large spin-down if the magnetic field of the star
is large, $B\sim 10^{13}$ G or greater. If the magnetic field is weaker, the
star enters this unstable region, and the r-modes grow until the heat released by
viscous dissipation is large enough to raise the temperature, leading the
star back to the stable region where it came from. Then, the r-modes are
damped until neutrino cooling moves the star back to the unstable region. The cycle
repeats until an equilibrium between amplitude-dependent heating and
neutrino cooling is achieved and the star stabilizes with a fixed amplitude
($\alpha\sim10^{-6}$) very close to the instability boundary (Fig. 3).
The star then evolves along the instability boundary, spinning down due
to the combination of magnetic dipole braking (which dominates for magnetic
fields larger than $\sim 10^9$ G, as inferred for ``classical'', young pulsars)
and gravitational radiation, providing a persistent source of
graviational radiation for $\sim200\rm{yr}$. This is qualitatively similar to the
evolution found by
Andersson et al. (2002) for strange stars, but with an important difference.
Since those authors do not consider reheating nor a magnetic field, the timescale
for the evolution along the instability boundary is set by the star's cooling
process rather than the magnetic torque.

\section{Conclusions}

We have found that the non-linear bulk viscosity terms cause no
dramatic changes in the evolution of a young neutron star.
They can decrease the maximum amplitude achieved by the r-modes, but
cannot completely saturate the modes.
The reheating makes the rapid spin-down phase take place at higher temperatures,
possibly simplifiying the models by avoiding the formation of a solid crust
and a superfluid core. The most important consequence
of reheating is that, with strong bulk viscosity at sufficiently low temperatures,
it can balance neutrino cooling, keeping the star close to the instability boundary,
and turning it into a persistent source of gravitational radiation for hundreds of years.

\acknowledgements

The possible importance of the nonlinear terms in the bulk viscosity was realized,
independently of us, by P. Arras, and that of reheating through bulk viscosity by Y. Levin.
Our work on this topic was initiated during A. R.'s participation in the
program on ``Spin and Magnetism in Young Neutron Stars'' at the ITP in
Santa Barbara, organized by L. Bildsten. A. R. thanks many of the participants of that
program, but specially G. Ushomirsky and S. Morsink, for teaching him
about r-modes and the CFS instability. We are also grateful to L. Lindblom
for information about his work, and the organizers of the Marsala workshop for
their efforts, which made an excellent conference happen. Most of the work
presented here was funded by FONDECYT grant 1020840.

\end{document}